\documentclass[seceq,preprint]{ptptex}

\usepackage{epsfig}
\notypesetlogo                       %comment in if to eliminate PTPTeX 
\preprintnumber{%<-- [..]: optional width of preprint # column.
UT-12-04\\IPMU 12-0019}
\def\lsim{\mathrel{\raise.3ex\hbox{$<$\kern-.75em\lower1ex\hbox{$\sim$}}}}
\def\gsim{\mathrel{\raise.3ex\hbox{$>$\kern-.75em\lower1ex\hbox{$\sim$}}}}
\def\stau{\widetilde{\tau}}

\title{%        %You can use \\ for explicit line-break
Stau-catalyzed $d$-$t$ Nuclear Fusion}

\author{%       %Use \scshape  for the family name
Koichi HAMAGUCHI$^{1,2}$, Tetsuo HATSUDA$^{1,3}$, Masayasu KAMIMURA$^{3}$
and Tsutomu T. YANAGIDA$^{1,2}$
}

\inst{%     %Affiliation, neglected when [addenda] or [errata]
$^{1}$Department of Physics, The University of Tokyo, 
Tokyo 113-0033, Japan\\
$^{2}$Institute for the Physics and Mathematics of the Universe, \\
 The University of Tokyo,  Chiba 277-8568, Japan\\
$^{3}$Theoretical Research Division, Nishina Center, RIKEN, \\
 Wako, Saitama 351-0198, Japan
}
\abst{%         %this abstract is neglected when [addenda] or [errata]
The gravitino of mass 10-100 GeV is a well motivated scenario in
supergravity. If the stau is the next lightest supersymmetry particle,
its life-time becomes order of $10^{6-8}$ sec. If it is the case the stau
makes a big impact on the nuclear fusion, since it is a charged
particle. In this paper we perform a detailed calculation of a
stau-catalyzed $d$-$t$ fusion.
We find that if certain technical conditions are satisfied,
 it is not hopeless to use the nuclear
fusion as a source of energy.
}

\begin{document}

\maketitle

\section{Introduction}

The gravitino mass $m_{3/2}$ is one of the most important parameter in supergravity, since it determines
the scale of supersymmetry (SUSY) breaking. From the phenomenological point of view, the gravitino 
mass can be in the range from 1 eV to 100 TeV. The gravitino of mass $O(10)$ GeV is particularly 
interesting, since it can be a dominant component of the DM in the universe~\cite{Bolz:1998ek}
and the thermal leptogenesis~\cite{Fukugita:1986hr} becomes consistent with cosmology if 
$m_{3/2}\simeq 10-100$ GeV. If it is indeed the case, the next lightest SUSY particle (NLSP) 
has a long lifetime of order $10^{6-8}$ sec.
The most natural candidate of the NLSP is the bino-like neutralino or the scalar partner of the tau lepton stau ($\stau$).\footnote{The above scenario is well realized in gauge mediation of SUSY breaking. It is remarkable that the Higgs boson mass around 125 GeV suggested in the recent report of the LHC experiments~\cite{ATLAS-CONF-2011-163,HIG-11-032} can be easily explained if one
considers the Higgs-messenger mixing~\cite{GMSB_125Higgs_1} or extra matters~\cite{GMSB_125Higgs_2} in gauge mediation models.
The latest LHC result has placed a lower bound on the stau mass as $m_{\stau}\gsim 221$~GeV~\cite{CMS_stau}.
}
If the $\stau$ is the NLSP and it has a long lifetime, it may 
give  impact on the big bang nucleosynthesis and   the nuclear fusion
as a charged catalyst.\footnote{The  stau decay may also destroy the success of BBN.
The cosmological problems caused by the catalysis, and by the stau decay, 
can be solved by a late-time entropy production~\cite{Buchmuller:2006tt}
or an enhanced coupling of stau to Higgs boson~\cite{stauHiggs} in the SUSY
standard model.}  Effect of such negatively-charged and long-lived massive particle 
for the production of light elements in the big bang has been extensively studied 
 (see e.g. the review, Ref.\citen{Iocco:2008va}.)  Also,
 three of the present authors  (K.H, T.H. and T.T.Y.) have discussed 
 the effects of
  long-lived $\stau$ on the $d$-$d$ nuclear fusion as a possible energy source~\cite{Hamaguchi:2006vp}.  Advantages as well as (serious) problems of
 the $d$-$d$ and $d$-$t$ fusion catalyzed by massive charged particles 
 have been also discussed before  \cite{Zweig,Okun} in analogy with the 
  muon catalyzed fusion.

 The purpose of this paper is to present a detailed quantum three-body 
   calculation of the $\stau$-catalyzed $d$-$t$ nuclear fusion,
   $\stau + d + t \to \alpha + n  +\stau $,
  especially  its  fusion rate
  and the sticking probability. Although it is certainly necessary
  to develop new technology to make such catalyzed fusion as a new source of energy,
  the present calculation would give one of the basis for such development.

Note that the following discussion is model-independent and applicable to any heavy charged particle with a sufficiently long lifetime. For instance, charged Wino NLSP can have a long lifetime if its mass is degenerate with the neutral Wino LSP. (See Ref.~\citen{Fairbairn:2006gg} for candidates for charged massive particles in various particle physics models beyond the Standard Model.) Thus, the ``stau'' in the following analysis can be replaced with any long-lived charged particle.

\section{Outline of the stau-catalyzed fusion in D-T mixture}

Let us first outline  how the $d$-$t$ fusion catalyzed by $\stau$
 proceeds inside the D-T mixture with a wide range of temperature
 10 K $\lsim T \lsim$ 1000 K.
 Suppose that a free $\stau$ is stopped in the D-T mixture, and
 the formation of the 1s state of the $\stau t$ and 
 $\stau d$ atom occurs
 through the capture process to the higher orbit,
$\stau + (te^-) \rightarrow (\stau t) + e^-$ and  $\stau + (de^-) \rightarrow (\stau d) + e^-$,
 followed by the de-excitation to their 1s states.\footnote{
 Note that there is no three-body bound state of $(\stau d t)$, 
 which we have checked explicitly by a three-body calculation.
 This is different from the case of $\stau$-$d$-$d$ system  discussed in Ref.
 \citen{Hamaguchi:2006vp}. Stau-catalyzed fusions in $\stau$-$d$-$d$ system will be 
 briefly discussed in Sec.~\ref{sec:results}.}

The basic  $d$-$t$ fusion reaction is
\begin{equation}
d + t \to \alpha + n  \quad(Q= 17.6 \,\,{\rm MeV}).
\label{eq:dt}
\end{equation}
The reaction rate of this process is  enhanced  
by the charge neutral $\stau t$ and $\stau d$ bound states 
in D-T mixture through the catalyzed processes
\begin{eqnarray}
({\stau}t)_{1s} + d &\to& \,\alpha + n + \stau , 
%\;\stau\alpha  + n 
\label{eq:dtX1}
\\
({\stau}d)_{1s} + t &\to& \,\alpha + n + \stau .
%\;\stau\alpha  + n 
\label{eq:dtX2}
\end{eqnarray}
Essential mechanism of the $\stau$ calalyzed fusion is similar
 to the muon catalyzed fusion:
 the long range Coulomb barrier is screened by the formation of 
 neutral bound states  tabulated in Table \ref{tab:bound-state}. 
Moreover, due to the large $\stau$ mass ($>$ 100 GeV) and small size (15 fm or less)
of $\stau t$ and $\stau d$ , the short range 
Coulomb repulsion is screened and even turns into 
an attraction which does not happen in  the case of $\mu$-catalyzed fusion
where  $\mu t$ and $\mu d$ have a large size of about 250 fm. 
We will show in the later sections by the fully quantum mechanical three-body
 calculations that the reaction rates
of (\ref{eq:dtX1}) and (\ref{eq:dtX2}) are enhanced
and the averaged reaction rate  $\lambda_{\rm f}$
 becomes $ 2.6 \times 10^8$ s$^{-1}$ independent
 of the temperature of the D-T mixture under consideration.

In the above reactions, there is always a possibility of 
forming  $\stau \alpha$ bound states in the final state. 
 Once such a sticking process takes place, sticked
  $\stau$ can no longer be used as a catalyzer, so that
the catalyzed fusion process is eventually stopped by the sticking.
 We calculate  this sticking probability to be
   $\bar{\omega}_{\rm s} =1.5 \times 10^{-3}$ as shown later, so that
 the energy product per $\stau$ is estimated as  
 $Q/\bar{\omega}_{\rm s} \simeq 12 $ GeV.

% XXX A brief discussion on the advantage of inflight d-t over in-flight dd here, XXX

\begin{table}[t]
\begin{center} 
\begin{tabular}{cccc}
\hline
System & $\mu_{r}$ & $a_B$ & $E_b$ 
\\ \hline \hline
$\mu d$   & 0.1~GeV    & 270~fm     & 2.7~keV\\
$\mu t$   & 0.1~GeV    & 270~fm     & 2.7~keV\\
$\stau d$ & 1.9~GeV & $15$~fm & $50$~keV\\
$\stau t$ & 2.8~GeV    & 10~fm     & 74~keV
\\ \hline
\end{tabular}
\end{center}
\caption{Reduced mass $\mu_{r}$, Bohr radius $a_B$ and binding energy
of the 1s state for muonic and stau atoms (for stau mass 200 GeV).
}
\label{tab:bound-state}
\end{table}

\section{Model and method}

Before entering the stau-catalyzed
three-body reactions (\ref{eq:dtX1}) and 
(\ref{eq:dtX2}),
we first calculate the cross section of the 
standard reaction (\ref{eq:dt})
by taking the same three-body calculation method of  
Refs.~\citen{Hamaguchi2007,Kamimura09} 
in which the fully-quantum method was applied  systematically to
the various types of stau-catalyzed big-bang nucleosynthesis reactions.
We explicitly follow \S 4 of Ref.~\citen{Kamimura09} on three-body
breakup reactions.  
We do not explicitly treat 
the complicated channel coupling between the
entrance and exit channels. Instead, we employ
an alternative model which is easy to incorporate into the
calculation of the three-body 
processes (\ref{eq:dtX1}) and (\ref{eq:dtX2}).
Namely, we take into account
the entrance $d$-$t$ channel alone and introduce a complex potential
$V_{d{\mbox -}t}(r)$ between $d$ and $t$:
\begin{equation}
V^{\rm nucl}_{d{\mbox -}t}(r) = V^{\rm (real)}_{d{\mbox -}t}(r) 
+ i V^{\rm (imag)}_{d{\mbox -}t}(r)   
\label{eq:vnucl}
\end{equation}
as seen in nuclear optical-model potentials.
Since there are no other open channel  
than the entrance and exit channels in (\ref{eq:dt})
at the energies concerned here,
the absorption cross section due to the imaginary potential represents 
the reaction cross section to the exit $\alpha+n$ channel.
We determine the potential $V^{\rm nucl}_{d{\mbox -}t}(r)$ so as to
reproduce  the observed cross section ($S$-factor) of 
(\ref{eq:dt}) at low energies ($\lsim$ 1 MeV).

Next, we incorporate this complex potential 
$V^{\rm nucl}_{d{\mbox -}t}(r)$ 
into the three-body Hamiltonian of the $d + t + \stau$ 
system (Fig.~1) with obvious notation,
\begin{eqnarray}
H=\! -{\hbar^2 \over 2m_{c}} \nabla^{2}_{{\boldsymbol r}_{c}}
        \! -{\hbar^2 \over 2M_{c}} \nabla^{2}_{{\boldsymbol R}_{c}}
      \!   + \!V^{\rm coul}_{t{\mbox -}\stau}(r_1) 
      \!   +\! V^{\rm coul}_{d{\mbox -}\stau}(r_2)
       \!  + \!V^{\rm coul}_{d{\mbox -}t}(r_3) 
       \!  + \!V^{\rm nucl}_{d{\mbox -}t}(r_3) 
       , \;\;    
\label{eq:ham}
\end{eqnarray}
where the kinetic-energy operator is equivalent for $c=1-3$.
We solve the Schr{\" o}dinger equation
$( H - E_{\rm total} )\, \Psi_{JM}=0$ for
the elastic scattering between $\stau t$ and $d$ 
and that between $\stau d$ and $t$. 
The absorption cross section obtained 
in this scattering calculation is considered to give the 
reaction cross section of the stau-catalyzed reactions
(\ref{eq:dtX1}) and (\ref{eq:dtX2}).

%%%%%%%%%%%%%%%%%  Fig. 2 ( Jacobi ) %%%%%%%%%%%
\begin{figure}[bth]
\begin{center}
\epsfig{file=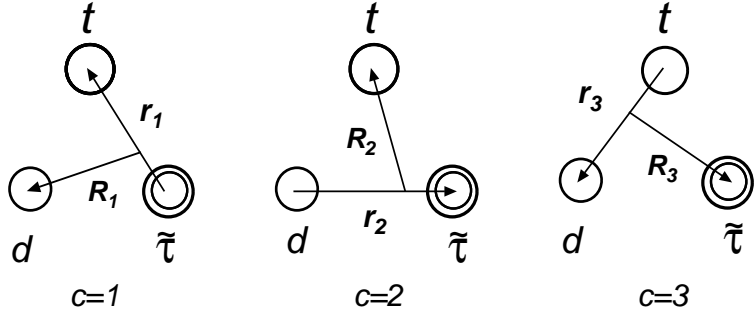,width=10cm,height=4.25cm}
\end{center}
\caption[]{ 
Three sets of Jacobi coordinates in
the $ d+t+\stau$ system. The scattering 
channels $\stau t + d$ and $\stau d + t$  are described  
by using the coordinate systems  $c=1$ and $c=2$, respectively. 
The coordinate $c=3$ is responsible for describing
the strong nuclear correlation and the nuclear fusion.
}
\label{fig:jacobi}
\end{figure}

%%%%%%%%%%%%%%%%%%%%%%%%%%%%%%%%%%%%%%%%%

In the following, we explain our method taking the case of
$\stau t+d$ reaction on $c=1$ (similarly for the 
$\stau d+t$ on $c=2$).
The total wave function with the angular momentum $J$ and
its $z$-component $M$ is written  as 
\begin{equation}
\Psi_{JM}=\phi^{(1)}_{1s}({\boldsymbol r}_1)\,
\chi^{(1)}_{JM}({\boldsymbol R}_1)   
        +\Phi^{({\rm corr.})}_{JM}\;,
\label{eq:wf}
\end{equation}
where $\phi^{(1)}_{1s}({\boldsymbol r}_1)$ represents 
the $1s$ wave function of  the $\stau d$ atom and 
$\chi^{(1)}_{JM}({\boldsymbol R}_1)$ for the
$(\stau d)_{1s}+d$ scattering wave.  
The scattering boundary condition imposed on  
$\chi^{(1)}_{JM}({\boldsymbol R}_1) ( \equiv \chi^{(1)}_J(R_1)
Y_{J M}({\widehat {\boldsymbol R}}_1) )$
is given by 
\begin{eqnarray}
\lim_{R_{1}\to\infty} R_1 \chi^{(1)}_{J}(R_1) = 
u^{(-)}_{J}(k_{1},R_{1})
-S^{J}_{1 \to 1}u^{(+)}_{J}(k_{1},R_{1}) ,
\label{asym}
\end{eqnarray}
where $u^{(-)}$ and $u^{(+)}$ are the asymptotic incoming and outgoing
wave function, and $\hbar^2 k_1^2 / 2M_{1}=E_{\rm total}-
E_{1s}^{(1)} \equiv E $, $E_{1s}^{(1)}$ being the energy of 
$(\stau d)_{1s}$.

The second term of 
(\ref{eq:wf}), $\Phi^{({\rm corr.})}_{JM}$, is introduced to describe
the short-range correlation along the coordinate ${\bf r}_3$
due to the strong nuclear interaction 
$V^{\rm nucl}_{d{\mbox -}t}({\bf r}_3)$; the correlation amplitude
is not included in the first scattering term, but plays an 
important role in the $d$-$t$ fusion process.
Since $\Phi^{({\rm corr.})}_{JM}$ is asymptotically vanishing amplitude,
it is expanded in terms of 
three-body Gaussian basis functions \cite{Hiyama2003} as
\begin{eqnarray}
&& \Phi^{({\rm corr.})}_{JM}  =  \sum_{nl, NL}
 b_{nl, NL} \; \big[\phi_{nl}({\rm r}_3) \, 
\psi_{NL}({\rm R}_3)\big]_{JM} ,\\
\label{eq:phi-close}
%\end{eqnarray}
%\begin{eqnarray}
&&\phi_{nlm}({\boldsymbol r}) =
r^l\:e^{-(r/r_n)^2} \:Y_{lm}({\widehat {\boldsymbol r}}) , \quad
\psi_{NLM}({\boldsymbol R}) = 
R^L\:e^{-(R/R_N)^2} \:Y_{LM}({\widehat {\boldsymbol R}}) ,
\label{eq:3gaussa}
\end{eqnarray}
where the Gaussian ranges are postulated to
lie in a geometric progression: 
\begin{eqnarray}
r_n=r_{\rm min}\, a^{n-1} , \;  (n=1-n_{\rm max}) 
\quad R_N=R_{\rm min}\, A^{N-1} . \; (N=1-N_{\rm max})
\label{eq:3gauss}
\end{eqnarray}

After solving the Schr\"{o}dinger equation to determine 
$S^J_{1 \to 1}$ and $ b_{nl, NL}$,\cite{Hiyama2003,Kamimura09}
 we derive the absorption cross section by
\begin{equation}
\sigma(E) 
=\frac{\pi}{k_1^2}\sum_{J=0}^{\infty} (2J+1)
       (1- \bigl|S^J_{1 \to 1} \bigr|^2)  .
\label{eq:sig1}
\end{equation}
This absorption cross section
is alternatively  expressed as 
\begin{equation}
\sigma(E)=
\frac{-2}{\hbar v_1} \langle \Psi_{JM} \,|  V^{\rm (imag)}_{d-t}(r_3)
\,| \Psi_{JM} \,\rangle ,
\label{eq:sig2}
\end{equation}
where $v_1$ is the c.m. velocity of the incident channel $(c=1)$.
It is to be noted that the two equivalent
$\sigma (E)$ utilize information from quite different parts
of the three-body wave function $\Psi_{JM}$, 
namely, the information
from the asymptotic part along ${\bf R}_1$ in the former
expression and that from the internal part along  ${\bf r}_3$ in 
the latter.
Therefore, it is a severe test of the numerical accuracy of 
the three-body calculation
to examine the agreement between  the two types of
$\sigma (E)$.  In our calculation below, we obtained
a precise agreement between their numbers 
in four significant figures.
%, which demonstrates the high accuracy of  our
%three-body calculation.

The reaction rate $\langle \, \sigma v \,\rangle$ at  temperature $T$
is expressed as %(cf. Eq.~(4-44) of Ref.~\citen{Clayton1983})
\begin{eqnarray}
\langle \, \sigma v \,\rangle & =& 
% \left( \frac{8}{\pi M_1}\right)^{\frac{1}{2}} 
(8/\pi M_1)^{1/2} (kT)^{-3/2}
\int_0^\infty \,\sigma(E) \,
{\rm e}^{-E/kT} dE \:,
\label{eq:rate}
\end{eqnarray}
where  $k$ is the Boltzmann constant.
It is  noted here that if, as we shall meet below,
$\sigma (E) \propto  1/ \sqrt{E}, $  %%(1/v_1)$, 
then the rate $\langle \, \sigma v \,\rangle$ becomes 
independent of temperature $T$. 
%%%%%%%%%%%%%%%%%%%%%%%%%%%%%%%%%%%%
%%%%%%%%%%%%%%%%%%%%%%%%%%%%%%%%%%%%
\section{Calculated results}
\label{sec:results}
%%%%%%%%%%%%%%%%%%%%%%%%%%%%%%%%%%%%
%%%%%%%%%%%%%%%%%%%%%%%%%%%%%%%%%%%%

%%%%%%%%%%%%%%%%%%%%%%%%%%%%%%%%%%%%
\subsection{Interactions}
%%%%%%%%%%%%%%%%%%%%%%%%%%%%%%%%%%%%

As for the the complex potential $V^{\rm nucl}_{d{\mbox -}t}(r)$,
we  take the same one as used in the study of muon 
catalyzed $d$-$t$ fusion in Ref.~\citen{Kamimura89},
where the fusion rate and the $\alpha \mu$ sticking
probability in the $dt\mu$ molecule were calculated using
$V^{\rm nucl}_{d{\mbox -}t}(r)$ having the Woods-Saxon shape
with five different types of parameter sets. 
Since the calculated results did not
significantly depend on the sets (all 
reproduce the cross section of (\ref{eq:dt}) very well for 
$E_{\rm c.m.}\lsim 1 $ MeV), 
we here employ a parameter set:
\{$V_0=-38.0$ MeV,  $R_0=$3.0 fm, $a_0=0.5$ fm\}  for the real part
and \{$W_0=-0.37$ MeV, $R_I=3.0$ fm, $a_I=0.5$ fm\} for the
imaginary part.  The Coulomb potential between $d$ and $t$
is constructed by assuming  
the Gaussian shape of the charge distribution of $d\, (t)$
which reproduces observed  r.m.s. radius.

%%%%%%%%%%%%%%%%%  Fig. 3 ( folding ) %%%%%%%%%%%
\begin{figure}[bth]
\begin{center}
\epsfig{file=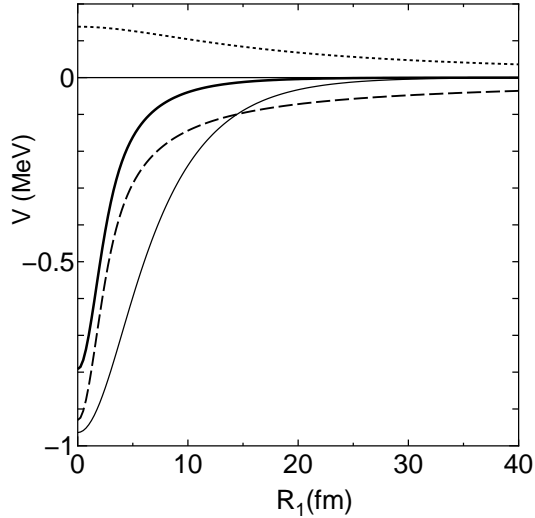,width=7cm,height=7cm}
\end{center}
\caption[]{ 
Coulomb and nuclear potential between $(\stau t)_{1s}$ and $d$
as a function of the coordinate $R_1$ in Fig.~\ref{fig:jacobi} 
($m_{\tilde{\tau}} \gg m_t$).
They are obtained by folding the $d$-$t$ potential into the
triton density in the $(\stau d)_{1s}$
atom except for the Coulomb $\stau$-$d$ 
potential by the dashed line. 
The dotted line is the $d$-$t$ folded part of the 
Coulomb $\stau t$-$d$ potential.  
The thick solid line is sum of these  two Coulomb potentials.
The thin solid line is the real part of the 
folded nuclear $\stau t$-$d$ potential.
}
\label{fig:folding}
\end{figure}
%%%%%%%%%%%%%%%%%%%%%%%%%%%%%%%%%%%%%%%%%

Coulomb and nuclear potential between
$\stau t$ and $d$ (between  $\stau d$ and $t$) 
are obtained by folding 
the Coulomb and nuclear $d$-$t$ potentials into the 
triton (deuteron) density of the 
$1s$-state of the $\stau t$ $(\stau d)$ atom.
In Fig.~2, we illustrate the case of the $\stau t$-$d$ potential.
The Coulomb $\stau $-$d$ 
potential is given by the dashed line. 
The dotted line is the $d$-$t$ folded part of the 
Coulomb $\stau t$-$d$ potential.  
The thick solid line is sum of these  two Coulomb potentials, which
almost vanishes for $R_1 \gsim 20 \,$ fm.
The thin solid line is the real part of the 
folded nuclear $\stau t$-$d$ potential.
The imaginary part (not illustrated) 
has the same shape of the real part but is
$-0.01$ MeV at $R_1=0$.

The screened Coulomb potential (thick solid line) is
attractive everywhere. The folded nuclear potential is
more attractive than this Coulomb potential.
Moreover, $t$ in  $\stau t$ ($d$ in  $\stau d$)
is moving with an averaged kinetic energy of 72 keV (48 keV).
We can thus expect a
huge enhancement of the fusion reaction rate at low energies
compared with the bare $d$-$t$ fusion and
the muon-catalyzed $d$-$t$ fusion-in-flight of the
$d$-$t\mu$ and $t$-$d\mu$ systems.{\footnote %%%%%%%%%%%%%%%%%%%%%%%%
{In the $d+t+\mu$ system, the $d$-$t$ Coulomb barrier
at short range is not screened by muon (the muon Bohr radius is
some 250 fm); this can be understood, in Fig.~3, by changing the sign of
the two Coulomb potentials in the dashed and dotted lines.
Furthermore, the kinetic energy of $d$ in $d\mu$ ($t$ in $t\mu$) 
is negligibly small compared with that of 
$d$ in $\stau d$ ($t$ in $\stau t$).} %%%%%%%%%%%%%%%%%

Therefore, it is of particular importance to describe properly 
the short-range $d$-$t$ relative motion at the
moment of the fusion. This is not satisfactorily done by the
first scattering term of (\ref{eq:wf}) since it does not 
explicitly include the $d$-$t$ coordinate (${\bf r}_3$), 
but is well realized by the second term, $\Phi_{JM}^{({\rm corr.})}$, 
using the basis functions $\phi_{nl}({\bf r}_3)$ along the $d$-$t$ 
coordinate.
We took the Gaussian parameters as
$\{n_{\rm max}=10; r_{\rm min}=0.2\, {\rm fm},
r_{\rm max}=10.0 \,{\rm fm}\}$
for $\phi_{nl}({\bf r}_3)$ and   
$\{N_{\rm max}=15; R_{\rm min}=1.0\, {\rm fm},
R_{\rm min}=50.0 \,{\rm fm}\}$
for $\psi_{NL}({\bf R}_3)$.  But, we found in the actual calculations
below that $l=L=J=0$ is sufficient for the low
energies $(E \lsim 10 ^1$ eV) concerned in the present paper.

We assume infinitely heavy stau.
However, as studied in Ref.\citen{Kamimura09}, results of the
stau-catalyzed reaction are known to depend little (by a few percents)
on the mass of stau ($\gsim  100$  GeV) 
except for the case of specific resonant reactions.

%%%%%%%%%%%%%%%%%%%%%%%%%%%%%%%%%%%%%
\subsection{Fusion rate}
\label{sec:fusionrate}
%%%%%%%%%%%%%%%%%%%%%%%%%%%%%%%%%%%%%

We  calculated the reaction (absorption) cross section $\sigma (E)$
of the $\stau t+d$ reaction (\ref{eq:dtX1}) for the energy region of
$10^{-4}$ eV $\lsim E \lsim 10^{2}$ eV (we are particularly interested in
10 K $\lsim T \lsim$ 1000 K, namely,
$10^{-3}$ eV $\lsim E \lsim 10^{-1}$ eV). 
We found that contribution from 
$J>0$ is negligible and 
the cross section is well represented by
\begin{equation}
\sigma_{\stau t+d}(E)= 
1.2 \times 10^{-20}/ \sqrt{E({\rm eV})} \;\;{\rm cm}^2 ,
\label{eq:sigxt}
\end{equation}
which follows the usual $1/v$-law for low energy reactions with a 
neutral-charge particle (here $\stau d$).\footnote{ %%%%%%%%%%%%%%%
If we omit, in (\ref{eq:wf}), the second term 
$\Phi^{({\rm corr.})}_{JM}$ for describing 
the strong $d$-$t$ nuclear correlation along ${\bf r}_3$,
the cross section becomes  smaller by one order.}   %%%%%%%%%%%%%%%%%%%%
%%%
With the $E$-dependence of the cross section in (\ref{eq:sigxt}), the rate  
$\langle \, \sigma v \,\rangle$ for the 
$\stau t+d$ reaction (\ref{eq:dtX1})
is expressed  independently of $T$, for $kT \lsim 10^1$ eV, as
\begin{equation}
\langle \, \sigma v \,\rangle_{\stau t+d}
= 9.7 \times 10^{-15} \;\; {\rm cm}^3
\,{\rm s}^{-1} .
\label{eq:ratext}
\end{equation}

Similarly, the reaction cross section of the
${\stau d+t}$ reaction (\ref{eq:dtX2}) is obtained as
\begin{equation}
\sigma_{\stau d+t}(E)= 
2.5 \times 10^{-21}/ \sqrt{E({\rm eV})} \;{\rm cm}^2
\label{eq:sigdt}
\end{equation}
for $E \lsim 10^{2}$ eV.
The reaction rate $\langle \, \sigma v \,\rangle_{\stau d+t}$ 
is expressed, for $kT \lsim 10^1$ eV, as\footnote{   %%%%%%%%%%%%%%%%%%%%%
If we employ another (longer-ranged) set of nuclear interaction 
in Ref.\citen{Kamimura09}, namely 
\{$V_0=-16.0$ MeV,  $R_0=$5.0 fm, $a_0=0.3$ fm\}  for the real part
and \{$W_0=-0.28$ MeV, $R_I=2.5$ fm, $a_I=0.3$ fm\} for the
imaginary part, we obtain
$\langle \, \sigma v \,\rangle_{\stau t+d}
= 2.0 \times 10^{-15} \;\; {\rm cm}^3
\,{\rm s}^{-1}$ and
$\langle \, \sigma v \,\rangle_{\stau d+t} 
= 9.5 \times 10^{-15} \;\; {\rm cm}^3
\,{\rm s}^{-1}$, which is close to the result in (\ref{eq:ratext}) and 
(\ref{eq:ratexd}), respectively.} 
%
%%%%%%%%%%%%%%%%%%%%%%%%%
\begin{equation}
\langle \, \sigma v \,\rangle_{\stau d+t}
= 2.4 \times 10^{-15} \;\; {\rm cm}^3
\,{\rm s}^{-1} .
\label{eq:ratexd}
\end{equation}

From the results (\ref{eq:ratext}) and (\ref{eq:ratexd}), 
we can say that, for temperature $T \lsim 10^5$ K,
the stau-catalyzed $d$-$t$ fusion
occurs with the {\it same} rate at any $T$.
For the density $n \simeq 4.25 \times 10^{22}$ atoms/cm$^3$,
irrespective of gas-, liquid- and solid-phases of the \mbox{D-T}~mixture,
the above reaction rates lead to
$\lambda_{\rm f} \simeq n \langle \, \sigma v \,\rangle
\simeq 2.6 \times 10^8$ s$^{-1}$, independently of temperature,
on the average of (\ref{eq:ratext}) and (\ref{eq:ratexd}).

The formation rate of the 1$s$ state of
the $\stau t \,(\stau d)$ atom in the D-T mixture 
%is the order of $10^{12}$~s$^{-1}$, which is much larger than the above fusion rate.
is expected to be much faster than the above fusion rate.
Then, the cycling rate, $\lambda_{\rm  c}$,
of the stau-catalyzed fusion is given by $\lambda_{\rm c}
\simeq 2.6 \times 10^8$ s$^{-1}$, which is compatible with the
the largest $\lambda_{\rm c}$ achieved so far in 
the muon-catalyzed $d$-$t$ fusion experiment.
The time scale of the one cycle is given by $1/\lambda_{\rm c}=
3.8 \times 10^{-9}$ sec.

%%%%%%%%%%%%%%%%%%%%%%%%%%%%
\subsection{Sticking probability}
%%%%%%%%%%%%%%%%%%%%%%%%%%%%

In the exit channel of the fusion reactions (\ref{eq:dtX1}) and 
(\ref{eq:dtX2}), some fraction of the $\alpha$ particles may 
be trapped by $\stau $
and form a  Coulomb bound state $\stau \alpha$,
though most of them escape into the $\stau \alpha$ continuum states.
If this $\stau \alpha$-sticking process happens, the fusion 
chain will be terminated. The probability
of sticking to the bound states gives a stringent
constraint on the number of fusion reactions per $\stau $.
For the muon-catalyzed fusion, the sticking probability is known to be
about 1\% (15\%) for $dt\mu \,(dd\mu)$.

In the sudden approximation where the instantaneous $dt$ fusion 
does not affect the states of infinitely heavy $\stau $, 
the sticking probability for the fusion-in-flight 
may be estimated 
from an overlap integral of the initial and final state wave functions
in the same manner of the sticking
after fusion in the $dt\mu\, (dd\mu)$ molecule:
\footnote{  %%%%%%%%%%%%%%%%%
For the fusion-in-flight, the sticking probability is
derived as follows according to the reaction theory 
(see Ref.~\citen{Kamimura86} for more details):
In the reaction (\ref{eq:dtX1}), we consider that
the $d$ and $\alpha$ are composed of
$n+p$ and $t+p$, respectively, and $p$ in $d$ is transferred to 
$t$ to form $\alpha$.
The transition matrix $T_{\rm f}$ to the final state f 
in the $\stau \alpha$ atom is exactly expressed 
with the obvious notation as
\vspace{-0.13cm}
\begin{equation}
T_{\rm f}=\langle \psi_\alpha({\bf r}_{tp}) \, 
\phi_{\rm f}({\bf r}_{\stau \alpha}) \,
{\rm exp}(i {\bf k}_{\rm f}\cdot {\bf r}_{\stau n}) \,
|\,V_{nt}({\bf r}_{nt})+ V_{np}({\bf r}_{np})\,|\,
\Psi^{\rm exact}_{\rm scatt.}\,
\rangle , \nonumber
\end{equation}
\vskip -0.13cm
\noindent
where  $\Psi^{\rm exact}_{\rm scatt.}$ is the exact wave function of
the reaction (\ref{eq:dtX1}).
If we assume $V_{nt}({\bf r}_{nt}) \propto \delta({\bf r}_{nt})$,
$V_{np}({\bf r}_{np}) \propto \delta({\bf r}_{np})$
and $\psi_\alpha({\bf r}_{tp})  \propto \delta({\bf r}_{tp})$,
which leads to ${\bf r}_{\alpha n}=0$ and 
${\bf r}_{\stau  n}={\bf r}_{\stau \alpha} (\equiv{\bf r})$,
and  replace  $\Psi^{\rm exact}_{\rm scatt.}$ by
our $\Psi_{JM}$ in (\ref{eq:wf}),  we then obtain
$T_{\rm f} \propto \langle \phi_{\rm f}({\bf r})\,
{\rm exp}(i {\bf k}_{\rm f}\cdot {\bf r}) \,|\,
\Phi_{\rm i}({\bf r})\,\rangle$, which is used in (\ref{eq:omega0f}).
}
%%%%%%%%%%%%% end of \footnote  %%%%%%%%%%
%
\begin{eqnarray}
&&\qquad \qquad
\omega_{\rm s} = \sum_{\rm f}^{\rm bound}\,\omega_{\rm s,\,{\rm f}}\:, 
\label{eq:omega0}
\\
&&\omega_{\rm s,\,{\rm f}}= \frac{ | \,\langle \phi_{\rm f}({\bf r})\,
{\rm exp}(i {\bf k}_{{\rm f}}\cdot {\bf r}) \,|\,
\Phi_{\rm i}({\bf r})\,\rangle \,|^2 } 
{\, |\langle \Phi_{\rm i}({\bf r})\,|\,\Phi_{\rm i}({\bf r})\,
\rangle\,|^2}\: ,
\label{eq:omega0f}
\end{eqnarray}
where $ \Phi_{\rm i}({\bf r})$ is the initial scattering wave function
(\ref{eq:wf}) at the instant of fusion $({\rm r}_3=0)$:
\begin{equation}
\Phi_{\rm i}({\bf r}) = \Psi_{JM}({\rm r}_3=0, {\bf R}_3 ={\bf r})
=\phi^{(1)}_{00}({\bf r})\,
\chi^{(1)}_{JM}({\bf r})   
        +\Phi^{({\rm corr.})}_{JM}({\bf r}_3=0,{\bf R}_3={\bf r})\;,
\label{eq:phi-i}
\end{equation}
and $\phi_{{\rm f}}({\bf r})$ denotes 
a normalized final state ($\stau  \alpha)_{\rm f}$
and ${\bf k}_{\rm f}$ is the wave vector
of  the relative motion between $n$ and $\stau \alpha$
determined by the relation $\hbar^2 k_{\rm f}^2/2m_n + E_{\rm f}=
17.6$ MeV. 
The summation in (\ref{eq:omega0}) is over all the {\it bound} 
states ($\stau  \alpha)_{\rm f}$ 
(we have $\omega_{\rm s}=1$ if the summation 
is taken over all the $\stau \alpha$
states including the continuum).

%%%%%%%%%%%%%%%%%%%%  Table 1: sticking 
\begin{table}[th]
\caption{Calculated $\stau \alpha$-sticking 
probability $\omega_{\rm s}$
and its partial components $\omega_{\rm s,\, {\rm f}}$.
}
\label{table:resonance}
\begin{center}
\begin{tabular}{cccccc} 
\hline \hline
\noalign{\vskip 0.1 true cm} 
 &  & $\quad$$\omega_{\rm s,\,{\rm f}}$ &   &   
&   $\omega_{\rm s}$  \\
\noalign{\vskip -0.15 true cm} 
&  \multispan4 {\hrulefill} &  
     \cr
\noalign{\vskip 0.05 true cm} 
reaction   
&  $1s$    
& $2s$   
& $2p$ 
&  others  
& total    \\
\noalign{\vskip 0.05 true cm} 
\hline 
\noalign{\vskip 0.15 true cm} 
 $\stau t+d$  
&  $1.1 \times 10^{-3}$  & $1.8 \times 10^{-4}$  
&  $4.1 \times 10^{-6}$ 
& $1.1 \times 10^{-4}$   
&  $1.4 \times 10^{-3}$      \\
\noalign{\vskip 0.2 true cm} 
 $\stau d+t$  
&  $1.4 \times 10^{-3}$  & $2.4 \times 10^{-4}$  
&  $5.3 \times 10^{-6}$ 
& $1.5 \times 10^{-4}$   
&  $1.8 \times 10^{-3}$      \\
%\vspace{-5 mm} \\
\noalign{\vskip 0.1 true cm} 
\hline\\
\end{tabular}
\end{center}
\end{table}
%%%%%%%%%%%%%%%%%%%%%%%%%%%%%%%%%%%%%%
%

Calculated $\stau \alpha$-sticking probability $\omega_{\rm s}$
and its partial components $\omega_{\rm s,\, {\rm f}}$
are listed in Table \ref{table:resonance} for the reactions (\ref{eq:dtX1}) 
and (\ref{eq:dtX2}).\footnote{  %%%%%%%%%%%%%%%%%%%%%%
If we employ another set of nuclear interaction that was
mentioned in one of the previous footnotes, 
we obtain the sticking probability
$\omega_{\rm s} \simeq 1.4 \times 10^{-3} \, 
(1.8 \times 10^{-3})$ for the $\stau t+d$ 
($\stau d+t)$ reaction, which is the same as
the result in Table \ref{table:resonance}.

Even if we omit, in (\ref{eq:wf}), the second term 
$\Phi^{({\rm corr.})}_{JM}$ for describing 
the strong $d$-$t$ nuclear correlation along ${\bf r}_3$,
the sticking probability $\omega_{\rm s,f}$ does not change
significantly since the second term works to increase much the
cross section but almost equally for all the final states.
} %%%%%%%%%%%%%%%%%%%%%%%%%%%%
If we average the two values of $\omega_{\rm s}$
taking the magnitude of the reaction rates (\ref{eq:ratext}) 
and (\ref{eq:ratexd}) as averaging weight, 
we have ${\bar \omega}_{\rm s}= 1.5 \times 10^{-3}$. This is several times
smaller than $\omega_{\rm s}(\simeq 0.01)$ 
in the muon-catalyzed $d$-$t$ fusion in the $dt\mu$ molecule.

The sticking probability 
listed in Table~\ref{table:resonance} does not depend on the
energy (for $E \lsim 10^{1}$ eV) concerned in this paper.
This is because of the following reason: 
Integration up to only $r \simeq 15$ fm
contributes to the numerator of Eq.~(\ref{eq:omega0f}).
In this region, as seen in Fig.~\ref{fig:folding},
the incident energy  is negligibly small
compared with the
depth of the attractive Coulomb and nuclear potentials.

Since  the number of fusion cycles available
before terminating due to the sticking loss of $\stau $
is given by $1/{\bar \omega_{\rm s}}$ (assuming no reactivation
of sticked $\stau $),
the energy product $E_{\stau dt}$ per $\stau $
is estimated as 
\begin{equation}
E_{\stau dt}=\frac{17.6  \;\;{\rm MeV}}{1.5 \times
10^{-3}} \simeq 12 \;\;{\rm  GeV} .
\label{eq:e_dtx}
\end{equation}

%%%%%%%%%%%%%%%%%%%
\subsection{stau-catalyzed $d$-$d$ fusion}

Here we shortly comment on the stau-catalyzed $d$-$d$ fusion.
It is much less effective than the stau-catalyzed $d$-$t$ fusion
studied above. The reason is as follows:
Differently from the $d$-$t$ case, the $\stau +d+d$
system has a bound state, the $(\stau dd)_{J=0}$ atom,
at 2.8 keV below the 
$(\stau d)_{1s}+d$ threshold. However, since there is no 
excited bound state, we cannot expect  a rapid formation of 
the $(\stau dd)_{J=0}$ atom via the Vesman's resonant
mechanism\cite{Vesman} (a bound state shallower than 
4.5 eV is required).
Moreover, the sticking probability after the fusion in the atom
is $\omega_{\rm s} \simeq 0.03$.
As for the stau-catalyzed $d$-$d$ fusion-in-flight,
the sticking probability is
$\omega_{\rm s} \simeq 0.02$, whereas 
the Q-value is about 3.7 MeV (average of 
$Q= 3.3~{\rm MeV}$ for
$d + d \to n+^3$He and 
$Q= 4.0~{\rm MeV}$ for
$d + d \to p+t$).
Thus, 
the energy product $E_{\stau dd}$ per $\stau $
is estimated to be
$E_{\stau dd} \simeq 3.7 \;{\rm MeV}/ 0.02 \simeq 
0.15 \,  {\rm GeV}$
(assuming no reactivation of sticked $\stau $),
which is much smaller than $E_{\stau dt} \simeq 12$ GeV.

%\pagebreak

%%%%%%%%%%%%%%%%%%%%%%%%%%%%%
\section{Discussion}
\label{sec:discussion}
%%%%%%%%%%%%%%%%%%%%%%%%%%%%%

Let us now briefly discuss a possible production of the staus in the laboratory.
We consider the  $\mu + N \ ({\rm nucleon})$  scattering with a fixed nuclear target.
The stau-production cross section depends on the spectrum of 
SUSY particles. 
In order to have an optimistic estimate of the number of produced stau,
let us estimate the slepton--production cross section by using the
sparticle production 
cross section in cosmic ray neutrino--nucleon scattering
studied e.g., in Ref.~\citen{Ahlers:2006pf}.
For the neutrino energy $E_\nu \simeq 2000$ TeV, 
for the SUSY model point SPS 7~\cite{Allanach:2002nj},\footnote{This model point is already excluded by the LHC~\cite{CMS-PAS-EXO-11-022}, but we use it just for illustration.}
the cross section is ${\cal O}(10^{-38}~{\rm cm^2})$~\cite{Ahlers:2006pf}.
Since all SUSY particles decay quickly to the staus, the 
stau-production cross section is also of ${\cal O}(10^{-38}~{\rm cm^2})$.
Thus, assuming a laboratory energy of the muon  $E_\mu \simeq {\cal O}(1000)$ TeV, 
and by further assuming a Fe target of ${\cal O}$ (km) length with the nucleon density $n_N\simeq 5\times 10^{24}/{\rm cm}^3$,
the number of produced staus per muon is estimated to be ${\cal O}(10^{-8})$.
This implies that we need at least $~10^8\times1000$ TeV$\, (10^{14}$ GeV) 
to produce a single stau.\footnote{This is, of course, an overoptimistic guesstimate. 
It will cost much more than 1000 TeV to create a 1000 TeV muon,
and it will not be easy to capture all the high-energy staus produced.}
On the other hand, one stau could reproduce $\simeq$12 GeV energy 
for a single
chain (namely, $1/\omega_{\rm s}$ cycles) of the $d$-$t$ fusion
as we have discussed in \S 3. Therefore, to make the present stau-catalyzed 
$d$-$t$ fusion to be of practical use, we need to recycle the stau 
at least $10^{13}$ times, 
even if the above optimistic estimate holds.
For the recycling, we should collect the 
inactive $\stau \alpha$ atoms and strip the $\alpha$ 
particle from the stau. It is beyond the scope of the present
paper to investigate possible reactivation mechanisms.

As shown in \S \ref{sec:fusionrate}, the time scale of the one cycle
of the stau-catalyzed $d$-$t$ fusion is $3.8 \times 10^{-9}$ sec.
This leads to the time scale of a single chain which is estimated as
$3.8 \times 10^{-9}\, {\rm sec} / (1.5 \times 10^{-3}) 
= 2.5 \times 10^{-6}$ sec.
Assuming that we find a sufficiently fast reactivation mechanism 
of stau, the lifetime of the stau should be longer than 
$2.5 \times 10^{-6} \,{\rm sec} \times 10^{13}=2.5 \times 10^7\, {\rm sec}
\simeq 300 \,{\rm days}$ at least for the output energy to exceed
the input energy. 
In order to make the present stau-catalyzed nuclear fusion 
an interesting source of energy, we thus need to find a
more efficient mechanism and/or technology for the stau production.

 Other than catalyzing the nuclear $d$-$t$ fusion, negatively charged $\stau$ may 
  also provide a new tool in nuclear physics.
  Indeed, if the $\stau$ is embedded in heavy nuclei, 
   it will form exotic Coulomb bound states with their level structures
    affected by the charge distribution of the nuclear interior. 
  Namely, the long-lived stau may be used as a probe of the deep interior of heavy nuclei.

\section*{Acknowledgement} 
This work was supported by Grand-in-Aid for Scientific research from
the Ministry of Education, Science, Sports, and Culture (MEXT), Japan,
No. 21740164 (K.H.), and No. 22244021 (T.T.Y. and K.H.).
T.H. is supported in part by MEXT Grant-in-Aid for Scientific Research
on Innovative Areas (No.20105003) and SPIRE (Strategic Program for Innovative
REsearch) Field 5.
This work was supported by World Premier International Research Center Initiative (WPI Initiative), MEXT, Japan.

\end{document}